\documentclass[12pt]{article}
\newcommand{\be}{\begin{equation}}
\newcommand{\ee}{\end{equation}}
\newcommand{\beas}{\begin{eqnarray*}}
\newcommand{\eeas}{\end{eqnarray*}}
\newcommand{\bea}{\begin{eqnarray}}
\newcommand{\eea}{\end{eqnarray}}
\newcommand{\ba}{\begin{array}}
\newcommand{\ea}{\end{array}}
\newcommand{\nn}{\nonumber}

\newcommand{\al}{\alpha}
\newcommand{\g}{\gamma}
\newcommand{\de}{\delta}
\newcommand{\s}{\sigma}

\newcommand{\si}{\sigma}
\begin{document}
\begin{titlepage}
\title{
{\bf  Gravity as the affine Goldstone  phenomenon and beyond}
}
\author{
Yu.\ F.\ Pirogov\thanks{E-mail: pirogov@mx.ihep.su}\\[0.5ex]
{\it Theory Division, Institute for High Energy Physics,  Protvino,}\\
{\it  RU-142281 Moscow Region, Russia}
}
\date{}
\maketitle
\abstract{\noindent 
The two-phase structure is imposed on the world continuum, with the
graviton emerging as the tensor Goldstone boson during the spontaneous
transition from the affinely connected phase to the metric one.
The physics principle of metarelativity, extending
the respective principle of  special relativity, is postulated.  
The theory of metagravitation  as the general nonlinear model
$GL(4,R)/SO(1,3)$ in the arbitrary  background continuum is built.
The concept of the Metauniverse as the ensemble of the regions of the
metric phase inside the affinely connected phase is introduced, and
the  possible bearing of the emerging multiple universes to the fine
tuning of our Universe is conjectured.}

\thispagestyle{empty}
\end{titlepage}

\section{Introduction}

The General Relativity (GR) is the well-stated theory attributing 
gravity to the Riemannian geometry of the space-time. Nevertheless,
the ultimate nature of gravity awaits, conceivably, its future
explanation. In this respect, it is of great interest
the approach to gravity  as the Goldstone phenomenon corresponding
to the broken global affine symmetry~\cite{Salam,Ogievetsky1}.
Originally, this approach was realized  as the nonlinear model
$GL(4,R)/SO(1,3)$  in the Minkowski background  space-time, 
as distinct from the geometrical framework of the~GR.

In the present paper, we adhere to the viewpoint that the above
construction is more than just the mathematical one, but has a
deeper physics foundation underlying it. In this respect, the  new
insights motivating and extending the
Goldstone approach to gravity  are put forward. 
Of principle, we go beyond the framework of the Riemannian geometry.
Namely, we start with the world continuum considered as the affinely
connected manifold without metric and end up in  the space-time
with the effective Riemannian geometry. 

Our main results are threefold:

({\em i\/})\  The  physics principle
of extended relativity is introduced as a substitution for that of
special relativity. It states the physics invariance, at an
underlying level, relative to the choice within the  extended set of
the local coordinates, including the inertial ones. The principle
justifies the pattern of the affine  symmetry breaking $GL(4,R)\to
SO(1,3)$ required for the Goldstone approach to gravity.  
 
({\em ii\/})\ The extended theory of gravity,
with the GR as the lowest approximation, is built as
the proper nonlinear model in an arbitrary background continu\-um. 
The natural hierarchy of the possible GR extensions,  according to
the accuracy of the affine symmetry realization, is put forward. 

({\em iii\/})\
The extended Universe, as the
ensemble of the Riemannian metric universes inside the affinely
connected world continuum, is considered. It is conjectured  that the
multiple universes may clarify the fine tuning problem of our
Universe.

The content of the paper is as follows. 
In sec.~\ref{aff}, the principle of extended relativity is introduced. 
The spontaneous breaking of the ensuing global symmetry, the affine
one, with the residual Poincare symmetry and the emerging tensor
Goldstone boson is then considered. In sec.~\ref{AGB}, the nonlinear
realization of the broken affine symmetry is studied.
In sec.~\ref{MN}, the respective nonlinear model 
in the tangent space is developed. Its prolongation to the 
space-time, the extended gravitation, is presented in
sec.~\ref{CC}. Finally, the concept of the extended Universe is
discussed in sec.~\ref{ESV}, with some remarks in conclusion.

\section{Metarelativity}\label{aff}

\subsection{Affine symmetry}

Conventionally, the GR starts by postulating that the world continuum,
i.e.\ the set of the world events (points),  is the Riemannian
manifold. In other words, a metric is imposed  on the world 
ab initio. The metric specifies all the fine properties of the
continuum converting the latter into the space-time. Nevertheless, not
all of the properties of the  space-time depend crucially on the
metric.\footnote{Cf.\ in this respect the reflections on the
space-time structure due
to E.~Schr\"{o}dinger~\cite{Schrodinger}.} To appreciate the deeper
meaning of the gravity  and the very space-time, one needs possibly go
beyond the Riemannian geometry. 

To this end, consider the space-time  not as 
a priori existing  but as emerging in the processes of the world
structure formation. Namely, suppose that at an underlying level the
continuum is endowed  only with the topological structure (without
metric, yet). More particularly, it  is the affinely connected
manifold. The affine connection supports  the detailed continuity
properties, such as the parallel transport of the vector fields, their
covariant derivatives, etc. In particular, the connection produces the
curvature tensor as a result of the parallel transport of a vector
around the infinitesimal closed contour. But there is yet no
geometrical structures which would be inherent in the metric, such as
the interval, distances, angles, etc. 

Let $x^\mu$, $\mu=0,\dots,3$  be the
world coordinates, generally, in the patches. There being, in absence
of the  metric, no  partition  of the continuum onto the space and
time, the index~$0$ has yet no particular meaning and
is just the notational one. In ignorance of the underlying
``dynamics'', consider all the structures related to the
underlying level of the world continuum as the background ones. Let
$\bar\phi^\lambda{}_{\mu\nu}(x)$ be the  background
affine connection and  let $\bar\xi^{\al}$, $\al=0,\dots,3$ be the
background affiliated coordinates where  the connection have a
particular, to be defined, form ${\bar
\phi}{}^{\g}{}_{\al\beta}(\bar\xi)$.\footnote{The
bar sign refers in what follows to the background. The indices $\al$,
$\beta$, etc., are those of the special coordinates, while the
indices $\lambda$, $\mu$, etc.,  are the arbitrary world ones.}  
The connections  are related as usually:
\be\label{eq:Gamma'}
{\bar \phi}{}^{\g}{}_{\al\beta}(\bar\xi)=
\frac{\partial x^\mu}{\partial \bar\xi^{\al}}
\frac{\partial x^\nu}{\partial \bar\xi^{\beta}}
\left(\frac{\partial \bar
\xi^{\g}}{\partial x^{\lambda}}\,
\bar\phi^{\lambda}{}_{\mu\nu}(x)
-\frac{\partial^2 \bar\xi^{\g}}{\partial x^\mu  \partial
x^\nu}\right).
\ee
It follows thereof, that the  parts antisymmetric and symmetric in
the lower indices  transform independently, respectively,
homogeneously and inhomogeneously. In particular,
being zero in a point in some coordinates, the antisymmetric  part
(the torsion) remains to be zero independent of the coordinates. Thus
one can adopt the background torsion to be absent identically. As for
the symmetric part, one is free to choose the special coordinates to
make the physics description as transparent as possible.
 
So, let $P$ be a fixed but otherwise arbitrary point (the reference
point) with  the world  coordinates $X^\mu$. 
Adjust to this point the local coordinates as follows:
\be\label{coordinates}
\bar\xi^{\al}=\bar\Xi^{\al}+\bar e^{\al}_\lambda(X)
\left((x-X)^\lambda+\frac{1}{2}
\bar\phi^\lambda{}_{\mu\nu}(X)\,(x-X)^\mu (x-X)^\nu\right)
+ {\cal O}((x-X)^3).
\ee
Here $\bar\Xi^{\al}\equiv \bar\xi^{\al}(X)$ and $\bar
e^{\al}_\lambda(X)\equiv \partial
\bar \xi^{\al}/\partial x^\lambda\vert_{x=X}$ is the tetrad, with 
$\bar e^{\lambda}_{\al}(X)$ being the inverse one. 
These parameters are still arbitrary and liable to further
specification. In the vicinity of~$P$, the affine connection now looks
like:
\be\label{eq:connection}
{\bar \phi}^{\g}{}_{\al\beta}(\bar\xi)= \frac{1}{2}
{\bar \rho}^{\g}{}_{\al\delta\beta}(\bar\Xi)\,
(\bar\xi-\bar\Xi)^{\de} + {\cal O}((\bar\xi-\bar\Xi)^2),
\ee
with ${\bar \rho}^{\g}{}_{\al\delta\beta}(\bar\Xi)$
being the background  curvature tensor in the reference point. 
In the coordinates chosen, the affine connection vanishes in the
reference point: $\bar \phi^{\g}{}_{\al\beta}(\bar\Xi)=0$. 

Now, consider the whole set of the local coordinates nullifying the
affine connection in the reference point $P$. Under the world
coordinates $x^\mu$ being fixed, one can choose a~priori any one of
the local coordinates $\bar\xi^\al$. The group of replacements
$\bar\xi^\al\to \bar\xi'^\al$ among the latter ones is
given by transformations:
\bea\label{Aa}
(A,a)\ : \ 
\bar\Xi^{\al}  & \to& \bar\Xi'^{\al} =
A^\al{}_\beta\bar\Xi^\beta+a^{\al},\nn\\
\bar e^{\al}_\mu&\to& \bar e'^{\al}_\mu=A^\al{}_\beta \bar
e^{\beta}_\mu,
\eea
with $A^\al{}_\beta$ being an arbitrary nondegenerate matrix and
$a^\al$ being an arbitrary  vector. The transformations
$(A,a)$ being independent of $\bar \xi$, the group is the global one.
This is the inhomogeneous general linear
group $IGL(4,R)=T_4 \odot GL(4,R)$ (the affine one).  Under these and
only under these transformations, the
affine connection remains still to be zero in the reference point. 
The respective coordinates will be called the local affine
ones.\footnote{Being understood, the term ``local'' will be
omitted in what follows.} In these coordinates, the world continuum is
approximated by  the affinely flat manifold in a neighbourhood of the
reference point $P$. Particularly, the
covariant derivative in the affine coordinates in the point $P$
coincides with the ordinary one.  Being changeable under $A$, the 
nonzero background torsion in the point $P$ would violate
explicitly the affine symmetry. Just to abandon this, the  torsion is
adopted to be zero identically. 

The affine group $IGL(4,R)$
is 20-parametric and extends the 10-parameter Poincare group
$ISO(1,3)=T_4 \odot SO(1,3)$  by the transformations  varying the
scales, the affine dilatations,\footnote{The world coordinates being
fixed, the affine dilatations  are to be distinguished from the
conventional ones in the space-time~\cite{Coleman}.} the rest being
the scale preserving, unimodular, affine transformations.  The
dilatations $A=\Delta$ belong  to the one-parametric multiplicative
group  of
the positive real numbers,
$\Delta^\al{}_\beta=e^{-\lambda}\delta^\al_\beta$,
with $\lambda$ any real. The unimodular affine
transformations are the 9-parametric part of  the special linear
group $SL(4,R)\ni A_0$, with $\mbox{det}A_0=1$ (times reflections). 

According to the well-known principle of special relativity,  the
present-day local physics laws  are unchanged under the
choice of the inertial coordinates, the Poincare group being the
physics invariance symmetry. Now, introduce the principle of
extended relativity, or {\it the metarelativity}, 
stating that the local physics laws at the
underlying level are  unchanged
relative to the choice of the affine coordinates. This extends the
physics invariance symmetry from the Poincare symmetry  to the affine
one.

\subsection{Spontaneous symmetry breaking}

Presently, there is known no exact affine symmetry.
Thus, the latter should be broken in transition from the
underlying level to the effective one. Postulate that this is
achieved due to the spontaneous emergence  of the background metric
$\bar \varphi_{\mu\nu}$ in the world continuum. The metric, 
with  the Minkowskian signature,  is assumed to be correlated with
the background affine connection so that to look in the affine
coordinates as
\be\label{eq:metric}
\bar \varphi_{\al\beta}(\bar\xi)=
\bar\eta_{\al\beta}-\frac{1}{2}
\bar
\rho_{\g\al\delta\beta}(\bar\Xi)\,
(\bar\xi-\bar\Xi)^{\g}(\bar\xi-\bar\Xi)^{\de}+{\cal
O}((\bar\xi-\bar\Xi)^3).
\ee
Here one puts   $\bar\eta_{\al\beta}\equiv \bar
\varphi_{\al\beta}(\bar\Xi)$ and 
$\bar \rho_{\g\al\delta\beta}(\bar\Xi)=
\bar \eta_{\g\g'}\,
\bar \rho^{\g'}{}_{\al\delta\beta}(\bar\Xi)$.
The metric eq.~(\ref{eq:metric}) is such that the Christoffel
connection $\bar\chi^\g{}_{\al\beta}(\varphi)$, determined by the
metric, matches with the  affine
connection $\bar\phi^\g{}_{\al\beta}$ in the sense, that the
connections coincide locally, up to the first derivative:
$\bar\chi^\g{}_{\al\beta}=\bar\phi^\g{}_{\al\beta}+{\cal
O}((\bar\xi-\bar\Xi)^2)$. 
This is quite reminiscent of the well-known fact that the metric in
the Riemannian manifold may be  approximated locally, up to the
first derivative, by the Euclidean metric. In the wake of the
emerging background  metric, there appears the (yet primordial)
partition of the world continuum onto the space and~time. 
 
Under the linearly realized affine symmetry, the  background metric
ceases to be invariant. But it still possesses an
invariance subgroup. To find it note that, without any loss of
generality, one can choose among the  affine
coordinates the particular ones with $\bar \eta_{\al\beta}$
being  the Minkowski tensor $\eta=\mbox{diag}\,(1,-1,-1,-1)$. The
respective coordinates will be called the background inertial
ones.\footnote{The latter ones are to be distinguished  from
the effective inertial coordinates, to be itroduced.}   
Under the affine transformations, one~has 
\be\label{eq:eta_lin}
(A,a)\ : \ \eta\to \eta'=A^{-1T}  \eta A^{-1}\ne\eta, 
\ee
whereas  the Poincare transformations still leave $\eta$
invariant:
\be
(\Lambda,a)\ : \   \eta\to \eta'= \Lambda^{-1T}  \eta
\Lambda^{-1}=  \eta.
\ee
It follows that the group of invariance  is isomorphous to the
Poincare group $ISO(1,3)\subset IGL(4,R)$ for any fixed $\bar
\eta_{\al\beta}$. Physically, the spontaneous symmetry
breaking corresponds to fixing, modulo the Poincare
transformations, the class of the distinguished coordinates among the
affine ones. These coordinates correspond to the particular  choice
for $\bar \eta_{\al\beta}$. Of course, the fact that the distinguished
coordinates are precisely those with the Minkowskian $\eta_{\al\beta}$
is no more than the matter of convention, corresponding to the proper
inner automorphism of the affine group.

Thus under the appearance of the metric, the affine symmetry is broken
spontaneously to  the residual Poincare one:
\be
IGL(4,R)\stackrel{M_A\ }{\longrightarrow} ISO(1,3).
\ee 
For the symmetry breaking scale $M_A$, one expects  a priori
$M_A\sim M_{Pl}$,  with $M_{Pl}$ being the Planck
mass. More particularly the relation between the scales
is discussed in sec.~\ref{CC}.
Due to the spontaneous breaking, the affine symmetry should be
realized in the nonlinear manner~\cite{NL}, with the nonlinearity
scale $M_A$, the residual Poincare symmetry being still realized
linearly. The unitary linear representations of the latter correspond
to the matter particles, as usually. The translation subgroup being
intact, the broken part coincides with $GL(4,R)/SO(1,3)$. The latter
should be realized in the Nambu-Goldstone mode. Accompanying  the
spontaneous
emergence of the metric, there should appear the 10-component affine 
Goldstone boson which corresponds to  the ten  generators of the
broken affine transformations. The effective field theory of
the Goldstone boson is given by  the relevant nonlinear model, to be
studied in what follows.

\subsection{Lorentz symmetry}

The group $GL(4,R)$  possesses the 16 generators
$\Sigma^{\al}{}_\beta$.
By means of $\eta_{\al\beta}$, one can redefine  the 
generators as
$\label{eq:gen} \Sigma^{\al\beta}\equiv \Sigma^{\al}{}_
\g \eta^{\g\beta}$ and substitute the later ones  by the symmetric and
antisymmetric combinations
$\Sigma_{\pm}^{\al\beta}=\Sigma^{\al\beta}\pm \Sigma^{\beta\al}$.
Clearly, this partition is affine noncovariant. 
The respective commutation relations read as follows:
\bea
\frac{1}{i}[\Sigma_{\pm}^{\al\beta},\Sigma_{\pm}^{\g\delta}]&=&
\eta^{\alpha \g}\Sigma_{-}^{\beta\delta}\pm
\eta^{\alpha\delta }\Sigma_{-}^{\beta\g}\pm (\alpha\leftrightarrow
\beta),\nn\\
\frac{1}{i}[{} \Sigma_{-}^{\al\beta},\Sigma_{+}^{\g\delta}]
&=&
\eta^{\alpha \g}\Sigma_{+}^{\beta\delta}+
\eta^{\alpha\delta }\Sigma_{+}^{\beta\g}- (\alpha\leftrightarrow
\beta).
\eea
The generators $\Sigma_-^{\al\beta}$ correspond  to the 
residual Lorentz symmetry, whereas  $\Sigma_+^{\al\beta}$  to the 
broken part of the affine symmetry. The broken generators  contain, in
turn, the dilatation one $i\eta_{\al\beta}\Sigma_+^{\al\beta}$. 
The latter commutes with all the generators and  is thus proportional
to unity in any irreducible representation. For the generators
$\sigma^\al{}_\beta$ in the adjoint
representation  one has
$(\sigma^\al{}_\beta)^\g{}_\delta=1/i\,
\delta^\al_\delta\delta^\g_\beta$, so that the respective generators
$\s_\pm^{\al\beta}$  are as follows:
\be \label{eq:adj}
(\s_\pm^{\al\beta})^\g{}_\delta=\frac{1}{i}(\de^\al_\de\eta^{\beta\g}
\pm \de^\beta_\de\eta^{\al\g}).
\ee

The above partition of generators is used in what follows in
constructing the nonlinear model. First, we study the three kinds of
the substance, i.e., the
affine Goldstone boson, matter and radiation,
which are characterized by the three distinct types of the nonlinear
realization. With these building blocks, we then construct the
nonlinear model itself.

\section{Nonlinear realization}\label{AGB}

\subsection{Affine Goldstone boson}

Let $\bar\xi^{\al}$ be the background inertial coordinates
adjusted to the space-time point~$P$. Attach to this  point 
the auxiliary linear space~$T$, the tangent space in the point.
By definition, $T$ is isomorphous to the Minkowski
space-time. The tangent space is the
structure space of the theory, whereupon the
realizations of the physics space-time
symmetries, the affine and the Poincare ones, are defined. 
Introduce in~$T$ the coordinates $\xi^{\al}$,
the counterpart of the background inertial
coordinates $\bar\xi^{\al}$ in the space-time. 
By construction, the connection in the tangent space is zero
identically. For the connection  in
the space-time  point~$P$ in the coordinates $\bar\xi^\al$ to be zero,
too,  the coordinates in the vicinity of the reference point are
to be related  as $\xi^{\al}= \bar\xi^{\al} +
{\cal O}((\bar\xi-\bar\Xi)^3)$.
The coordinates $\xi^\al$ are the ones, wherein  
all the constructions in~$T$ are originally built. The latter ones
being done, one can use in $T$ the arbitrary coordinates.

According to ref.~\cite{NL}, the  nonlinear realization of the
symmetry~$G$ spontaneously
broken to the symmetry $H\subset G$ can be built on the
quotient space $K=G/H$, the residual subgroup $H$ serving as the
classification group. In the present case, one is interested in the  
pattern $GL(4,R)/SO(1,3)$, with the quotient space consisting of all
the broken  affine transformations.
Let $\ae(\xi)\in K$ be the coset function on the tangent space. It can
be represented by a group element $k(\xi)\in G$.
Under the affine transformations $(A,a)$,
the representative group element is to transform in the vicinity of
the reference point~as
\be\label{eq: k}
(A,a)\ :\  k(\xi) \to \ k'(\xi')=
A k(\xi)\Lambda^{-1},
\ee
where $\Lambda$ is the appropriate element of the residual
group, here the Lorentz one $SO(1,3)$.
One has similarly: $k^{-1}\to \Lambda k^{-1}A^{-1}$. 
In the same time, by the very  construction, the Minkowskian~$\eta$
stays invariant under the nonlinearly realized affine symmetry: 
\be\label{eq:eta}
(A,a)\ : \ 
\eta\to\eta'=\Lambda^{-1T}\eta\Lambda^{-1}=\eta,
\ee
in distinction with the linear realization, eq.~(\ref{eq:eta_lin}).
Accounting for eq.~(\ref{eq:eta}), one gets in the other terms:
\be\label{eq:k}
(A,a)\ :\  k(\xi)\eta \to
\ k'(\xi')\eta=
A k(\xi)\eta\Lambda^{T}.
\ee

To represent unambiguously the coset by the  group element $k$, one
should impose on the latter some auxiliary condition. E.g., require
$k$ to
be pseudosymmetric in the
sense that $k\eta=(k\eta)^T$ (and similarly for $k^{-1}$). 
This ensures  that $k$ has ten independent components, indeed,  
in accord with the ten broken generators.
Under the affine transformations, this results in the restriction  
$Ak\eta\Lambda^{T}=\Lambda\eta k^T A^T$.
This entails implicitly the  dependence of the
Lorentz transformation on the Goldstone boson:
$\Lambda=\Lambda(A,k)$. Hereof, the term ``nonlinear'' follows.
This construction implements the nonlinear 
realization of the whole broken group  $GL(4,R)$, 
the residual Lorentz subgroup  $SO(1,3)$ being still realized
linearly, i.e., $\Lambda(A,k)\vert_{A=\Lambda}\equiv \Lambda$. And
what is more, the dilatations~$\Delta$ being Abelian, one gets
$\Lambda(\Delta,k)= 1$, so that $\Lambda(A,k) =\Lambda(A_0,k)$, with
$A\equiv \pm  A_0\Delta$ and $A_0 \in SL(4,R)$.

By doing as above, one retains only the independent Goldstone
components but looses the local Lorentz symmetry.\footnote{Such
a procedure was adopted in refs.~\cite{Salam, Ogievetsky1}.} For this
reason, we will not impose any auxiliary condition. Instead, we 
extend the affine symmetry by the hidden local symmetry\footnote{The
hat sign refers in what follows to the
local Lorentz symmetry.} 
$\hat H\simeq H=SO(1,3)$, with the symmetry breaking pattern $G\times
\hat H\to H$.  For quantities in the tangent space, one
should distinguish now two types  of indices: the affine ones, acted
on by the global affine transformations~$A\in G$, and the Lorentz
ones, acted on by the local Lorentz transformations~$\Lambda(\xi)\in
\hat H$. To make this difference explicit, designate the affine
indices in the tangent
space as before $\al, \beta$, etc, while the Lorentz ones as $a, b$,
etc. The Goldstone field is represented in this case by the arbitrary
$4\times 4$ matrix ${\kappa}^\al{}_a(\xi)$ (respectively,  
${\kappa}^{-1}{}^a{}_\al$), which  transforms  similar to $k$  by 
eq.~(\ref{eq: k}) but with arbitrary~$\Lambda(\xi)$. In what
follows, it is understood that the Lorentz indices are manipulated by
means of the Minkowskian $\eta_{ab}$ (respectively, $\eta^{ab}$). 
So, in the component notation, 
$\kappa\eta$ looks like $\kappa^{\al a}$
(similarly, $\eta \kappa^{-1}$ is~$\kappa^{-1}{}_{a\alpha}$).
This is the linearization of the nonlinear model, with the
extra Goldstone degrees of freedom being unphysical due to  the 
gauge Lorentz transformations~$\Lambda(\xi)$.
The auxiliary gauge boson, corresponding to the
generators $\hat\Sigma^{ab}_-$ of the local Lorentz
symmetry, is expressed due to the equation of motion through 
$\kappa$ and its derivative. With this in mind, the abrupt
expressions entirely in terms of $\kappa$ and its
derivative are used in what follows. The versions
differ by the higher order corrections.

\subsection{Matter}

The affine symmetry contains the
Abelian, though broken,  subgroup of the affine dilatations. For this
reason, the generic matter fields $\phi$ may additionally be
classified by their affine scale dimension $l_\phi$,\footnote{The
latter is to be distinguished from the conventional scale dimension.}
so that:
\be\label{eq:phi}
(A,a)\ :\
\phi(\xi)\to\phi'(\xi')=e^{l_\phi\lambda}\hat\rho_\phi(\Lambda)
\phi(\xi),
\ee
with $\hat\rho_\phi(\Lambda)$ taken in the proper Lorentz
representation. According to eq.~(\ref{eq: k}), the scale dimension of
$\kappa$ is $l_\kappa=-1$ (respectively, $l_{\kappa^{-1}}=1$).
Thus, with  account for the transformation $\mbox{\rm det}\, \kappa\to
e^{-4\lambda} \mbox{\rm det}\, \kappa$ under dilatations, one can
rescale the matter fields to the effective ones 
$\hat\phi=(\mbox{det\,}\kappa)^{ l_\phi/4 }\phi$.
The new fields are affine scale invariant, i.e.\ correspond to  
$ l_{\hat\phi}=0$, and transform simply as the local Lorentz
representations. They are to be
used in constructing the nonlinear model. If the
affine symmetry is not explicitly violated, only the rescaled  matter
fields  enter the action. In any case, one can choose $\hat\phi$ and
$\mbox{det\,}\kappa$ as the independent field variables. 
Thus, instead of $\hat\phi$, the matter fields 
will be designated in what follows simply as $\phi$ with $l_\phi=0$.

\subsection{Radiation}

From the point of view of the nonlinear realization, the gauge bosons
of the internal symmetries constitute one more separate kind of the
substance, the radiation. 
By definition, the gauge boson fields  $V_{\al}$ transform under the
affine transformations linearly as the derivative
$\partial_{\al}\equiv\partial/\partial
\xi^{\al}$:
\be
(A,a) \ : \ V(\xi)\to V'(\xi')=A^{-1T} V(\xi),
\ee
corresponding thus to the scale dimension $l_V=1$.
For this reason, redefine the gauge fields as
$\hat V_a =\kappa^{\al}{}_a V_{\al}$.
The new fields transform as the local Lorentz vectors
\be\label{eq:barV}
\hat V(\xi)\to  \hat V'(\xi')=\Lambda^{-1T}  \hat V(\xi)
\ee
and correspond to  $\hat l_{ V}=0$. These gauge fields are
to be used in the model building. Altogether, this exhausts
the description  of all the three kinds of the  substance: the
affine Goldstone boson, matter and radiation.

\section{Nonlinear model}\label{MN}

\subsection{Nonlinear connection}

To explicitly account  for both the affine symmetry and the local
Lorentz one, it is convenient to start with the composite objects
transforming only under the latter symmetry. Clearly, any nontrivial
combinations of
$\kappa$ and $\kappa^{-1}$ alone transform explicitly under $A$. Thus
the derivative terms are inevitable. To describe the latter ones,
introduce the Maurer-Cartan one-form:
\be\label{eq:Maurer}
\hat\Omega=\eta \kappa^{-1}d \kappa,
\ee
with  $ d \kappa$ being the ordinary differential of~$\kappa$. 
Under the affine transformations $\kappa\to \kappa'=Ak\Lambda^{-1}$,
the one-form transforms as the local Lorentz representation:
\be\label{eq:Delta_trans}
\hat\Omega(\xi) \to
\hat\Omega'(\xi')=\Lambda^{-1T}\hat\Omega(\xi)\Lambda^{-1}+ 
\Lambda^{-1T}\eta d\Lambda^{-1},
\ee
with $d \Lambda$ being the ordinary differential of $\Lambda(\xi)$.
Here use is made of the relation $\eta\Lambda\eta=\Lambda^{-1T}$ for
the Lorentz transformations. 

In the component notation, the so defined one-form looks like
$\hat\Omega_{ab}$. Decompose it as 
\be
\hat\Omega_{ab}\equiv \sum_\pm \hat\Omega^\pm_{ab}=\sum_\pm 
[\eta \kappa^{-1}d\kappa]^\pm_{ab},
\ee
where $[\dots]^\pm$ means the  symmetric and antisymmetric parts,
respectively. 
One sees that $\hat\Omega^\pm_{ab}$ transform independently of each
other:
\be\label{eq:MC}
\hat\Omega^\pm(\xi)\to 
\hat\Omega'^{\pm}(\xi')=\Lambda^{-1T}\hat\Omega^{\pm}(\xi) 
\Lambda^{-1} +\delta^\pm,
\ee
where
\bea
\delta^-&=&\Lambda^{-1T}\eta d\Lambda^{-1},\nn\\
\delta^+&=&0.
\eea
Transforming homogeneously, the symmetric part
$\hat\Omega^+$ can naturally be associated with the nonlinear
covariant
differential of the Goldstone field. At the same time, the
antisymmetric part
$\hat\Omega^-$ transforms inhomogeneously and allows one to define the
nonlinear covariant differential of the matter fields:
$\label{eq:DPhi} D\phi=(d +
i/2\,\hat\Omega^{-}_{ab}\,\hat\Sigma^{ab}_\phi)\phi$,
with $\hat\Sigma^{ab}_\phi$ being the Lorentz generators in the
representation $\hat\rho_\phi$. The so defined
$D \phi$ transform homogeneously, like $\phi$ themselves. 

The generic nonlinear
covariant derivative $ D_{\al}\equiv D/ d \xi^{\al}$
transforms as the affine vector. The effective covariant derivative, 
which  transforms as the local Lorentz vector, can be constructed as
follows: 
\begin{equation}\label{eq:covdir}
\hat D_a\equiv \kappa^{\al}{}_a  D_{\al}
= \kappa^{\al}{}_a  D/d \xi^{\al}.
\end{equation}
Thus  one gets for the covariant derivative of the one-form:
\be\label{one-form}
\label{eq:B}
\hat\Omega^{\pm}_{abc}=
\kappa^{\g}{}_c \,\hat\Omega^\pm_{ab}/ d\xi^{\g}= [\eta \kappa^{-1}
\hat\partial_c^{\phantom \pm}\! \kappa]^{\pm}_{ab},
\ee
where  
\be
\hat\partial_c \equiv
\kappa^{\g}{}_c \partial_{\g}= \kappa^{\g}{}_c \partial/\partial
\xi^{\g}
\ee
is the effective, Goldstone boson dependent, partial derivative. 
It follows that $\hat\Omega^{-}_{abc}$ could be used as the 
connection for the nonlinear realization.
Note, that this expression precisely corresponds to the case 
of  the nonlinear realization of the spontaneously broken internal
symmetry, where this connection is determined uniquely. 
But in the present case of the space-time
symmetry, the coordinates transform under the same group as the
fields. This results in the possible ambiguity of the nonlinear
connection. 

Namely, the transformation properties of the
covariant derivative do not change if one adds to the
above minimal connection the properly modified
terms $\hat\Omega^+_{abc}$, the latter ones
transforming homogeneously. For reason justified later on in this
section, we choose for the nonminimal connection the following special
combination: 
\bea\label{eq:c}
\hat\omega_{abc}&=&\hat\Omega^-_{abc} +
\hat\Omega^{+}_{cab} -\hat\Omega^{+}_{cba}\nn\\
&=& [\eta \kappa^{-1}\hat\partial_c^{\phantom \pm}\! \kappa]^-_{ab} +
[\eta \kappa^{-1}\hat\partial_b^{\phantom \pm}\! \kappa]^+_{ca}-
[\eta \kappa^{-1}\hat\partial_a^{\phantom \pm}\! \kappa]^+_{cb}.
\eea
The nonlinear covariant derivative of the matter fields now
becomes
\be\label{eq:DA'}
\hat D_c \phi= 
\Big(\hat\partial_c +
\frac{i}{2}\,\hat\omega_{abc}\,\hat\Sigma_\phi^{ab}\Big)\,\phi.
\ee
$\hat D_c \phi$ transforms homogeneously and
can be used in model building.

\subsection{Gauge interactions}

\paragraph{Internal symmetry}

Let  $\hat V_a$ be the  generator valued gauge fields for the
internal gauge  symmetry.
The gauge fields are  supposed to be coupled universally
via  the nonlinear connection, eq.~(\ref{eq:c}).
With account for the Lorentz generators $-(\hat\sigma_-^{ab})^T$
in the contravariant adjoint representation  (being given by
eq.~(\ref{eq:adj}) with the obvious substitution for the indices), one
gets for the nonlinear derivative of the fields:
\be\label{eq:barDc} 
\hat D_a\hat V_b=\Big( \delta^c_b\hat\partial_a+\hat \omega^c{}_{ba}
\Big)\hat V_c.
\ee
It follows thereof, in particular, that  $\hat D_c\eta_{ab}=0$. Define
the gauge strength as usually:
\be\label{eq:barFca}
\hat F_{ab}
=\Big(\hat D_a +i\hat V_a\Big) \hat V_b-(a\leftrightarrow b).
\ee
It follows that the so defined gauge strength takes the form 
$\hat F_{ab}=\kappa^\al{}_a \kappa^\beta{}_b F_{\al\beta}$, with 
\be\label{eq:F_albe}
F_{\al\beta}=\Big(\partial_\al +iV_\al\Big)V_\beta-
(\al\leftrightarrow \beta).
\ee
Thus  $\hat F_{ab}$ does  possess  the correct
transformation properties with respect to both the affine symmetry and
the internal gauge symmetry.

\paragraph{Lorentz symmetry} 

Further, consider the local Lorentz symmetry as the gauge one with the
connection $\hat \omega_c\equiv 1/2\,
\hat\omega_{abc}\,\hat\Sigma^{ab}_-$, where  $\hat\Sigma^{ab}_-$ are
some generic Lorentz generators.  Define
the corresponding  gauge strength  for the affine Goldstone boson as 
\be 
\hat G_{cd}=(\hat\partial_c+i\,\hat\omega_c)\,\hat \omega_d
 -(c\leftrightarrow d)
\equiv \frac{1}{2}\hat R_{abcd}\hat\Sigma^{ab}_-.
\ee
This gives
\be
\hat R_{abcd}=
\hat\partial_{c}\, {\hat \omega}_{abd}-
{\hat \omega}^f{}_{ac}\,{\hat \omega}_{fbd}
-(c\leftrightarrow d).
\ee
This quantity  transforms homogeneously as the local Lorentz
tensor (and similarly for its partial contraction $\hat
R_{bd}\equiv\hat R^a{}_{bad}$). The total contraction 
\be
\hat R\equiv \hat R^{ab}{}_{ab}=2\hat\partial_a\hat
\omega^{ab}{}_b-
\hat \omega^{fa}{}_a\hat\omega_f{}^b{}_b+
\hat\omega^{fab} \hat\omega_{fba}
\ee
is the local Lorentz scalar and can be used in building the Lagrangian
for the Goldstone boson.

\subsection{Lagrangian}

\paragraph{Lorentz invariant form}

The constructed objects can serve
as the building blocks for the  nonlinear model
$GL(4,R)/SO(1,3)$ in the tangent space. Postulate the
equivalence principle in the sense that the  tangent space Lagrangian
should not depend explicitly on the tangent space counterpart of the
background curvature
$\bar\rho_{\g\al\de\beta}$, eq.~(\ref{eq:metric}).
Thus, the Lagrangian  may  be written as the general  Lorentz  
invariant function built of $\hat R$, 
$\hat F_{ab}$, $\hat D_a \phi$ and $\phi$ themselves. As usually, we
restrict ourselves by the terms containing two derivatives at the
most.

The generic Lorentz (and thus affine) invariant Lagrangian in the
tangent space is 
\be \label{eq:L_Lor}
L=L_{g}(\hat R) + L_{r}(\hat
F_{ab})
+  L_m(\hat D_a \phi,\phi).
\ee 
In the above, the basic Goldstone  Lagrangian $L_{g}$ is as
follows:
\be\label{eq:L_g}
L_{g}= c_g M_A^2\left(-\frac{1}{2}\, 
\hat R(\hat\omega_{abc})+\Lambda\right),
\ee
with $c_g$ being a dimensionless constant, to be chosen, and $\Lambda$
proving to be in what follows the cosmological constant. The
radiation Lagrangian $L_{r}$ is as usually   
\be
L_{r}=-\frac{1}{4}\, \mbox{tr}(\hat F^{ab}\hat F_{ab}).
\ee
Finally,  $L_m$ is the proper matter Lagrangian.\footnote{The matter
Lagrangian is normalized so that
$L_m\vert_{\phi=0}= 0$.} As for the radiation
and matter, their Lagrangian could
well be the affine invariant Lagrangian of the Standard Model or of
any its extension. In fact, the given nonlinear model can accomodate
any field~theory.

\paragraph{Affine invariant form}

The Lagrangian above gives the basic dynamical description
of the affine Goldstone boson, radiation and  matter. The local
Lorentz representations are necessary to construct the Lagrangian. The
latter being built, one can rewrite it in terms of the
affine representations. This allows one  to make explicit the
geometrical
structure of the theory and to relate it with the gravity. This is
achieved by the proper regrouping  the factors  $\kappa^\al{}_a$
and $\kappa^{-1}{}^a{}_\al$, so that to make the
affine indices to be explicit. The Lagrangian now becomes
\be\label{eq:L}
L= c_g M_A^2\left(-\frac{1}{2}\, R (\g_{\al\beta})
+\Lambda\right)  +
{L}_{r}(F_{\al\beta})
+  { L}_m (D_{\al}\phi,\phi ).
\ee 
Here 
\be\label{eq:gamma'}
\g_{\al\beta}=\kappa^{-1}{}^a{}_{\al}\eta_{ab}\kappa^{-1}{}^b{}_{\beta
}
\ee
transforms as the affine tensor 
\be\label{eq:g_aff}
(A,a)\  :\ \g_{\al\beta}\to \g'_{\al\beta}=A^{-1}{}^\g{}_\al
\g_{\g\delta}A^{-1}{}^\delta{}_{\beta}.
\ee
It proves that $ R(\g_{\al\beta})=\hat R(\hat\omega_{abc})$ can
be expressed as the contraction $R=R^{\al\beta}{}_{\al\beta}$ of the
tensor $R^\g{}_{\al\delta\beta}\equiv \kappa^{\g c}
\kappa^{-1}{}^a{}_\al
\kappa^{-1}{}^d{}_\de  \kappa^{-1}{}^b{}_\beta \hat R_{cadb}$, the
latter in turn being related with $\g_{\al\beta}$ as the
Riemann-Christoffel curvature tensor with the metric. In this, all the
contractions of the affine indices are understood with 
$\g_{\al\beta}$ (respectively, $\g^{\al\beta}$). 

Similarly,  $ D_{\al}\phi$ looks like the generally  covariant
derivative for the  matter fields: 
\be
D_{\g}\phi=\Big(\partial_\g+\frac{i}{2}\,\omega_{ab\g}\hat\Sigma^{ab}_
\phi
\Big)\phi,
\ee
with the spin-connection 
\be
\omega_{ab\g}\equiv\kappa^{-1}{}^{c}{}_{\g}\,
\hat\omega_{abc} = 
\kappa^\beta{}_a \nabla_{\g} \kappa^{-1}{}_{b\beta} -(a\leftrightarrow
b).
\ee
In the above, $\nabla_{\g}
\kappa^{-1}{}_{b\beta}\equiv(\de^\al_\beta\partial_\g
- \Gamma^\al{}_{\beta\g}) \kappa^{-1}{}_{b\al}$ is
the covariant derivative 
calculated with the Christoffel connection 
\bea\label{eq:Christoffel}
\Gamma^{\al}{}_{\beta\g}&=&
\kappa^{\al a}
\kappa^{-1}{}^{b}{}_\beta \kappa^{-1}{}^{c}{}_\g\,\hat\omega_{abc}\,
+\kappa^\al{}_a\partial_\g \kappa^{-1}{}^a{}_\beta\nn\\
&=&\frac{1}{2}\,\g^{\al \de}\Big(\partial_{\beta}\g_{\de \g}+
\partial_{\g}\g{}_{\de\beta}
-\partial_{\de}\g{}_{\beta\g}\Big).
\eea
In particular, one gets $\nabla_\g\g_{\al\beta}=0$ as the
affine counterpart  of the Lorentz relation  
$\hat D_c\eta_{ab}=0$.
For the radiation Lagrangian, one has the usual expression
\be
L_r=-\frac{1}{4}\,\mbox{tr}(F^{\al\beta}F_{\al\beta}),
\ee
with $F_{\al\beta}$ given by eq.~(\ref{eq:F_albe}). Finally, the
matter Lagrangian is obtained straightforwardly from $L_m$, 
eq.~(\ref{eq:L_Lor}), with account for eq.~(\ref{eq:gamma'}) and the
relation $\hat D_a= \kappa^\al{}_a D_\al$, eq.~(\ref{eq:covdir}).

Clearly,  $L_g$ looks like the GR Lagrangian in the tangent space
considered as the effective\footnote{For short, the term ``effective''
will  be omitted, while that ``background'' will, in contrast, be
retained.} Riemannian manifold with 
the metric~$\g_{\al\beta}$, the  Christoffel connection
$\Gamma^{\g}{}_{\al\beta}$, the Riemann-Christoffel curvature tensor
$R^\g{}_{\al\delta\beta}$, the Ricci tensor~$R_{\al\beta}$, the
Ricci scalar~$R$ and the tetrad $\kappa^{-1a}{}_{\alpha}$
(the inverse one $\kappa^{\al}{}_a$). 
This is in no way accidental. Namely, as it is  shown in
ref.~\cite{Ogievetsky1}, under the special choice  of the
nonlinear connection eq.~(\ref{eq:c}),
the Lagrangian becomes conformally invariant, too. In this,  the
dilaton  of the conformal symmetry coincides with the affine
dilaton, while the vector Goldstone boson of the conformal symmetry,
proving to be  the derivative of the dilaton, is auxiliary. 
Further, according to the theorem due to
Ogievetsky~\cite{Ogievetsky2},
it follows that the theory which is invariant both under the conformal
symmetry and  the global affine one is generally covariant, as well.
After the proper choice of the metric, this imposes the effective
Riemannian structure onto the tangent space.
In the world coordinates, this will result in the generally covariant
theory (the GR, in particular).
Precisely this property justifies the special choice eq.~(\ref{eq:c})
for the nonlinear connection, with  the Goldstone boson being the
graviton in disguise.

\section{Metagravitation}\label{CC}

\subsection{General covariance}

Accept the tangent space Lagrangian  as that for the  space-time,
being valid in the background inertial coordinates in the
infinitesimal neighbourhood of the reference point~$P$.
After the subsequent multiplication  of the Lagrangian by the
covariant volume element $(-\g)^{1/2}\,d^4
\bar\Xi$\,, with  $\g\equiv \mbox{det}\g_{\al\beta}$, one gets
the contribution into the action of the infinitesimal
neighbourhood of the  point $P$. Now one has to convert this
contribution into the arbitrary world coordinates
and to sum over the whole space-time. 

The relation between the  background inertial  and  world coordinates
is achieved by means of the background  tetrad  
$\bar e^{\al}_\mu(X)$, eq.~(\ref{coordinates}), with the world
metric being as follows:
\be\label{eq:g}
 g_{\mu\nu}(X)= \bar e^{\al}_\mu(X) \g_{\al\beta}(\bar\Xi)
\bar e^{\beta}_\nu(X).
\ee
With account for eqs.~(\ref{Aa}), (\ref{eq:g_aff}), 
this metric is invariant under the  affine transformations
\be
(A,a)\  : \  g_{\mu\nu}\to g_{\mu\nu}.
\ee
By the very construction,  the world coordinates are unchanged as
well:
\be
(A,a)\ : \  X^\mu\to X^\mu.
\ee
As a result, the effective interval $d s^2= g_{\mu\nu} dX^\mu dX^\nu$
remains invariant, too.
 
Now, introduce the effective tetrad related with the background one
as 
\be\label{eq:bare}
e_\mu^a(X)=\kappa^{-1}{}^a{}_{\al}(\bar\Xi)\,\bar e_\mu^{\al}(X).
\ee
The effective tetrad transforms as  the local Lorentz vector:
\be\label{eq:Le}
e_\mu(X)\to e'_\mu(X)=\Lambda(X)\, e_\mu(X).
\ee
Due to the local Lorentz transformations $\Lambda(X)$, one can 
eliminate six components out of~$e_\mu^a$, the latter having 
thus ten independent components. In this terms, the world metric~is
\be
g_{\mu\nu}(X)=  e^a_\mu(X) \eta_{ab}
e^b_\nu(X).
\ee
In other words, the tetrad $e^a_\mu$ defines the effective inertial
coordinates. Physically, eq.~(\ref{eq:bare}) describes the
disorientation of the
effective inertial and background inertial frames depending on 
the distribution of the affine Goldstone boson (and thus the
gravity).

With account for the relation  $d \,\bar \Xi^{\al}=\bar
e^{\al}_\mu d X^\mu$
between the displacements of the point~$P$ in the background
inertial and  world coordinates, and thus
$\partial\,\bar\Xi^\al/\partial
X^\mu=\bar e^\al_\mu$, one has 
\be\label{eq:gamma}
\Gamma^{\lambda}{}_{\mu\nu}=\bar e^\lambda_{\al}\bar
e_\mu^{\beta}\bar e_\nu^{\g}
\Gamma^{\al}{}_{\beta\g}
+\bar e^\lambda_{\al}\partial_\mu\bar e_\nu^{\al},
\ee
where $\partial_\mu=\partial/\partial X^\mu$.
This  can be rewritten as usually:
\be
\Gamma^\lambda{}_{\mu\nu}=
\frac{1}{2}\,g^{\lambda\rho}\Big(\partial_{\mu}g_{\rho
\nu}+
\partial_{\nu}g_{\rho\mu}
-\partial_{\rho}g_{\mu\nu}\Big).
\ee
By construction, the world indices are manipulated via  $g_{\mu\nu}$
and $g^{\mu\nu}$.
The spin-connection looks in the world coordinates like
\be\label{eq:omega}
\omega_{ab\mu}=\omega_{ab\g} \bar e^\g_\mu=
e^\nu_a\nabla_\mu e_{ b\nu}-(a\leftrightarrow b),
\ee
with  the generally covariant derivative $\nabla _\mu$
defined via  the Christoffel connection
$\Gamma^\lambda{}_{\mu\nu}$, as usually. 
Respectively, the covariant derivative of the matter fields looks like
\be\label{eq:DPhi'}
{ D}_\mu \phi=\Big(\partial_\mu 
+\frac{i}{2}\,\omega_{ab\mu}\hat\Sigma_\phi^{ab}\Big) \phi.
\ee
In the similar way, 
one finds the  usual expressions for the Riemann-Christoffel 
tensor $R^\lambda{}_{\mu\rho\nu}(g)$,  the Ricci tensor $R_{\mu\nu}=
R^\lambda{}_{\mu\lambda\nu}$ and  the Ricci scalar
$R= g^{\mu\nu}R_{\mu\nu}$. The same is true for the gauge strength: 
\be
F_{\mu\nu}=(\partial_\mu +iV_\mu)V_\nu-(\mu\leftrightarrow \nu).
\ee 

Plugging the above modified objects into the Lagrangians for  the
affine Goldstone boson, radiation and matter and  integrating with the
invariant volume element one gets the total  action,  the
Einstein-Hilbert one including:
\be\label{eq:L_AGB'}
S=\int \left[
M_{Pl}^2\left(-\frac{1}{2}\, R(g_{\mu\nu})+\Lambda\right)
+L_r( F_{\mu\nu})
+L_m({D}_\mu \phi,\phi)\right]
(- g)^{1/2}\,d^4 X,
\ee
with $ g\equiv \mbox{det}\, g_{\mu\nu}$. In the above, the constant
$c_g$ in eq.~(\ref{eq:L}) is chosen so that $c_g M_A^2=1/(8\pi
G_N)\equiv M^2_{Pl}$, with $G_N$ being the Newton's constant and
$M_{Pl}$ being  the  Planck mass.
Varying the  action  with respect to the metric
$g^{\mu\nu}$ one arrives at the well-known equation of motion for
gravity:
\be\label{eq:Einstein'}
G_{\mu\nu}
=M_{Pl}^{-2}T_{\mu\nu}.
\ee
In the above, $G_{\mu\nu}$ is the gravity tensor:
\be\label{G_mu_nu}
G_{\mu\nu}\equiv {R}_{\mu\nu} -\frac{1}{2}\,R g_{\mu\nu}+\Lambda
g_{\mu\nu}
\ee
and $T_{\mu\nu}=T^{\,r}_{\mu\nu}+ T^{\,m}_{\mu\nu}$ is the
conventional
energy-momentum tensor of the radiation and  matter,  produced by 
$L_r$ and~$L_m$.

\subsection{General covariance violation}

By choosing  the generally covariant Lagrangian  in the
tangent space  one arrives at the generally covariant theory in the
space-time. Modulo the  choice of the Lagrangian, such a theory is
unique, independent of the choice of the coordinates. In particular,
one manages to express everything exclusively in the internal
dynamical terms (but for the numerical parameters). Under extension of
the tangent space Lagrangian beyond the generally
covariant one,  the theory in the space-time ceases to be generally
covariant and thus unique. It depends not only on the Lagrangian but
on the choice of the coordinates, too. Relative to  the general
coordinate transformations, the variety of  theories divides into
the observationally inequivalent classes, each of which  being
characterised by the particular set of the background
parameter-functions. Precisely the latter ones
make the coordinates distinguishable. A~priori,
no one of the sets of parameter-functions is preferable. Which one is
suitable (if any), should be determined by observations. 
Each class of theories consists of the equivalent
theories related by the residual covariance group. The latter
consists of the coordinate transformations leaving the background
parameter-functions invariant. On the contrary, one class can be
obtained from another
by the  coordinate transformations changing  these
parameter-functions. Weakening the requirements on the bundling of the
tangent spaces, one 
extends the set of the admissible theories, but arrives, instead,  at
the dependence of the theory in the space-time on the more elaborate
properties of the background. 

To clarify the corresponding parameter-functions,  construct the
background metric 
\be \label{barg}
\bar g_{\mu\nu}(X)=  \bar e^{A}_\mu(X) \eta_{AB}
\bar e^{B}_\nu(X),
\ee
with the inverse one 
\be\label{barg-1}
\bar g^{-1\mu\nu}(X)=\bar e^\mu_A(X) \eta^{AB}\bar e^\nu_B(X).
\ee
Here the generic index  $A$ means $a$  or $\al$,  as appropriate
(and similarly for $B$, $b$, $\beta$,  etc). This  metric
transforms intricately under the arbitrary affine transformations:
\be
(A,a)\ : \ \bar g_{\mu\nu}\to \bar g'_{\mu\nu}=\bar e_\mu^T A^{T}\eta
A
\bar e_\nu\neq \bar
g_{\mu\nu},
\ee 
though being invariant under the Poincare transformations
$(\Lambda,a)$. The metric $\bar g_{\mu\nu}$ is the next-of-kin to the
primordial one $\bar\varphi_{\mu\nu}$,
eq.~(\ref{eq:metric}).  The former  approximates the latter as
closely as possible in the lack of the knowledge of the primordial
background curvature $\bar\rho^\g{}_{\al\de\beta}$,
eq.~(\ref{eq:metric}).
According to the equivalence principle, this curvature does not enter
the tangent space Lagrangian and thus is inessential. Respectively,
the Christoffel connection
$\bar\Gamma^\lambda{}_{\mu\nu}(\bar g)$ approximates with the
same accuracy the  Christoffel connection
$\bar\chi^\lambda{}_{\mu\nu}(\bar
\varphi)$ and thus the primordial affine  connection
$\bar\phi^\lambda{}_{\mu\nu}$, i.e., 
$\bar\Gamma^\lambda{}_{\mu\nu}\simeq
\bar\chi^\lambda{}_{\mu\nu}\simeq \bar\phi^\lambda{}_{\mu\nu}$. 
So, in the reasonable assumptions, it suffices
to know only the background metric~$\bar g_{\mu\nu}$.

\paragraph{Affine symmetry preservation}

To be more specific, consider the  extension of the
tangent space Lagrangian for the Goldstone boson by means of
the terms depending explicitly on
$\hat\Omega^{+}{}_{abc}$, eq.~(\ref{one-form}).
E.g., one can add to the basic Goldstone  Lagrangian
eq.~(\ref{eq:L_g}) the quadratic piece:
\be
\Delta L_{g}^{(0)}=\frac{1}{2}\,\varepsilon_0 M_{Pl}^2\, 
\hat\Omega^{+b}{}_{ba}\hat\Omega^+{}_c{}^{ca}
\ee
where $\varepsilon_{0}$ is a dimensionless constant.
With account for eqs.~(\ref{eq:B}) and (\ref{eq:Christoffel}), one
gets for $\sigma_\al\equiv -\kappa^{-1}{}^a{}_\al{} 
\hat\Omega^{+b}{}_{ba}$ the relation
$\sigma_\al= \Gamma^\beta{}_{\beta\al}(\g)=  \partial_\al \s$, 
where $\s\equiv 1/2 \,\ln(-\g)$ and   $\g\equiv
\mbox{det}\,\g_{\al\beta} = -(\mbox{det}\,\kappa^a{}_\alpha)^{-2}$. 
In the affine terms, one has
\be
\Delta L_{g}^{(0)}=\frac{1}{2}\,\varepsilon_0 M_{Pl}^2\,
\gamma^{\al\beta}
\partial_\al\sigma
\partial_\beta \s.
\ee 
This Lagrangian violates the conformal symmetry in the tangent space
(more particularly, the local dilatation), as well as the general
covariance, though not violating the global affine symmetry.  

It follows from eq.~(\ref{eq:g}) that $ \g=-g/\bar g $, where 
$\bar g=\mbox{det\,}\bar  g_{\mu\nu}=-(\mbox{det\,} \bar
e^\al_\mu)^2$.
In the world coordinates, the Lagrangian $\Delta L_{g}^0$ becomes
\be
\Delta L_g^{(0)}=\frac{1}{2}\,\varepsilon_{0} M_{Pl}^2\,g^{\mu\nu}
\partial_\mu \sigma \partial_\nu \sigma ,
\ee
with 
\be
\sigma\equiv 1/2\, \ln (g/\bar g) 
\ee
and
\be
\partial_\mu \sigma=\Gamma^\lambda{}_{\lambda\mu}(g)-
\bar\Gamma^\lambda{}_{\lambda\mu}(\bar g).
\ee
Thus, all the  background dependence in the given case is determined
only  by the scalar density $\bar g$.
Note that $\partial_\mu \si$ transforms homogeneously  and thus can
not be eliminated by the 
coordinate transformations, though each one of the contributions  
could separately be nullified by the choice of 
coordinates.\footnote{Under  $\bar g=-1$, the given GR extension
reduces to that of ref.~\cite{Buchmuller}.}

Varying the total action (the Lagrangian $\Delta L_g^{(0)}$ included),
one arrives at the modification of the equation of motion for gravity,
eq.~(\ref{eq:Einstein'}), with the extra piece in the gravity tensor
$\Delta G^{(0)}_{\mu\nu}$.
Introducing the derivative couplings of $\si$ with matter, not
violating explicitly the affine symmetry, one would get the extra
piece $\Delta T^{\, m}_{\mu\nu}$ in the  energy-momentum tensor for
matter. Clearly, the  modified theory, though not
being generally covariant, is consistent with  the unimodular
covariance, i.e., that leaving $\bar g$ (as well as  $g$) invariant.
The unimodular covariance is  next-of-kin to the general one. 
Due to this residual covariance, the given GR extension describes
only three physical degrees of freedom corresponding to  the
``scalar'' and massless tensor gravitons. 
In the case $\varepsilon_{0}=0$, the general covariance is restored
eliminating thus one more degree of freedom. This  leaves just two of
them with helicities $\lambda=\pm 2$, as it should be for the massless
spin-2 particle.

The extra  terms in the Goldstone boson Lagrangian
would make physical the other latent degrees of freedom of the
gravity field, but by the cost of further violating the
general covariance. E.g., one could supplement the Goldstone
Lagrangian by the other independent quadratic pieces:
\bea
\Delta L_{g}^{(1)}&=&\varepsilon_1 M_{Pl}^2\, 
\hat\Omega^+{}_a{}_b{}^b \hat\Omega^+{}^{ac}{}_{c},\nn \\
\Delta L_{g}^{(2)}&=&
\varepsilon'_2 M_{Pl}^2\, \hat\Omega^+_{abc}
\hat\Omega^{+abc}+
\varepsilon''_2  M_{Pl}^2 \, \hat\Omega^+_{abc}
\hat\Omega^{+cab}.
\eea 
This  would, in particular, violate  causality for the ``vector''
graviton, as well as modify interactions for the tensor graviton.
Phenomenologically, these and similar modifications could be done as
small as necessary by the choice of the numerical parameters
$\varepsilon$. This is insured by the  fact that in the limit when
these parameters vanish, the symmetry of the theory increases up to
the general covariance.

\paragraph{Affine symmetry violation}

The derivative couplings above  preserve the affine symmetry, though
violating the general covariance. 
It is  conceivable another way of the general covariance
violation by introducing into  the tangent space Lagrangian the
potential $U_g(\kappa)$, which contains only the derivativeless
couplings of the Goldstone boson.  Of necessity, this would
explicitly violate the affine symmetry, too. To preserve 
the local Lorentz symmetry, the potential should depend only on
$\g_{\al\beta}$ (and/or~$\g^{\al\beta}$).
Not to violate the global Lorentz symmetry, too,  the potential is to
be chosen as a Lorentz invariant function as follows: 
\be
U_g=U_g\Big( \mbox{det\,} \gamma, 
\mbox{tr\,}(\gamma\eta)^n\Big),
\ee
with any degree $n$. In the above, one
puts $(\g\eta)_A{}^{B}\equiv \g_{AA'}\eta^{A'B}$, were as
before  $A=a$ or $\al$, etc, as appropriate. 
At $n<0$, one uses the relation
$(\gamma\eta)^{n}=(\eta\gamma^{-1})^{|n|}$,
with $\gamma^{-1\al\beta}\equiv\gamma^{\al\beta}$. 
It follows thereof that in the world terms the potential
should depend on $ g\bar g^{-1}$:
\be
U_g=U_g\Big( \mbox{det\,}( g\bar g^{-1}),\mbox{tr\,}
(g\bar g^{-1})^n\Big),
\ee
with the background metric  given by eqs.~(\ref{barg}),
(\ref{barg-1}). 
Generally, one has $\bar g^{-1\mu\nu}\neq \bar g^{\mu\nu}\equiv
g^{\mu\mu'} g^{\nu\nu'} \bar g_{\mu'\nu'}$ (and
similarly, $ \bar g^{-1}_{\mu\nu}\equiv g_{\mu\mu'} g_{\nu\nu'} 
\bar g^{-1 \mu'\nu'}
\neq \bar g_{\mu\nu})$).
At the negative $n$,  one puts 
$(g\bar g^{-1})^n =(\bar g g^{-1})^{|n|} $, with
$g^{-1\mu\nu}\equiv g^{\mu\nu}$. 
In this, the terms depending only on $\mbox{det\,}( g\bar
g^{-1})= e^{2\si}$ are unimodular covariant. 
The potential above corresponds to the case of the  most general
graviton mass with the Lorentz symmetry
preservation.\footnote{For the
theory of the massive tensor field in the
Minkowski background space-time see, e.g., ref.~\cite{Ogievetsky3}.
For the phenomenology of the graviton mass and for further references
on the subject cf., e.g., ref.~\cite{Visser}.} 

With advent of the potential, the only
modification of the gravity equation of motion  is the appearance of
the extra piece $\Delta G^{(U)}_{\mu\nu}$ in the l.h.s.\ of
eq.~(\ref{eq:Einstein'}).  The Bianchi identity states the covariant
divergenceless of the gravity tensor $G_{\mu\nu}$,
eq.~(\ref{G_mu_nu}).  Due to this identity, there appear four
constraints on the metric field and its first derivative.
These constraints substitute the Lorentz-Hilbert gauge condition. Thus
at the level of the  equation of motion, the theory describes
six physical degrees of freedom,  the massive scalar
and tensor gravitons. Choosing different contributions to $U_g$, one
can vary the relation between the respective 
masses. In the limit of vanishing  potential, the general
covariance is restored and one recovers smoothly the~GR with the
massless two-component tensor graviton.

One more similar source of the general covariance  violation could be
due to the derivativeless couplings of the affine Goldstone boson with
matter. Violating  the affine symmetry, all the derivativeless
couplings are expected
naturally to be suppressed (if any). This is in distinction with the
extra terms depending on the derivatives of the Goldstone
boson. The latter terms  also result in  the general covariance
violation. Nevertheless, being affine invariant, they are
not expected a priori to be small. 

This exhausts the foundations of the effective field theory of the
gravity, radiation and matter. The above theory, embodying the GR and
its extensions in the framework of the affine symmetry and the general
relativity, 
may be called {\it the metagravitation}.

\section{Metauniverse}\label{ESV}

\subsection{World continuum} 

The ultimate goal of the Goldstone approach to gravity
is to go beyond the effective metric theory and to build the
underlying premetric one. In what follows, we present some hints of
the respective scenario. Of necessity, we will be 
very concise, just to indicate the idea.

The forebear of the space-time
is supposed to be the world continuum. At the very least, the latter
is to  be endowed  with the defining structure, 
the continuity in the topological sense. 
Being covered additionally  with the patches of
the  smooth real coordinates~$x^\mu$, $\mu=0,1, \dots,$
$d-1$ (index 0 having yet no particular meaning),  the continuum
acquires the structure of the differentiable manifold of the
dimension $d$ (4, for definiteness). 
There exist in the continuum the tensor densities, in
particular, the volume element. Thus, the integration over the
manifold is allowed. 
But this does not suffice to define the  covariant derivative and thus
to get the covariant differential equations, etc. 
Suppose now, that the continuum can exist in two phases with the
following affinity properties.

\paragraph{Affine connection}

Being endowed with the primordial affine connection 
$\bar\phi^\lambda{}_{\mu\nu}$, the continuum becomes the affinely
connected manifold. Generally, the connection is the 64-parametric
structure. It defines the parallel transport of the world vectors, as
well as their covariant derivatives. The parallel transport along the
infinitesimal closed
contour defines, in turn, the background curvature tensor 
$\bar\rho^\lambda{}_{\mu\rho\nu}$ and thus its contraction
$\bar\rho_{\mu\nu}=\bar\rho^\lambda{}_{\mu\lambda\nu}$ (but not  yet
the scalar~$\bar\rho$). To every point $P$, there can be attached  the
coordinates~$\bar\xi^{\al}$,  where the symmetric part of the
connection locally
nullifies, the manifold becoming thus locally affinely flat. 
This defines the global affine symmetry. For the symmetry
to be exact, the antisymmetric part of the connection, the torsion,
should be trivial, with the connection being just 40-parametric.  
In this phase, there is yet no metric and thus  no space and
time directions, even no definite space-time signature, no lengths
and angles, no preferred Lorentz group and thus
no finite dimensional spinors, no preferred Poincare group and thus 
no conventional particles, no invariant
intervals, no quadratic invariants, no causality, etc.
Though there can be implemented the principle of the least action with
the primitive Lagrangians,
the world structure is still rather dull. Nevertheless, it should
ultimately lead to the spontaneous transition from the given phase
to the metric~one.

\paragraph{Metric}

Further, being endowed spontaneously with the 
metric $\bar \varphi_{\mu\nu}$ having the Minkows\-kian
signature, the continuum becomes the metric space, i.e., the
space-time. The metric is much more restrictive 10-parametric
structure. It defines
the background  Riemannian geometry. Accompanying  the emergence of
the metric and the spontaneous breaking of the affine symmetry,  
there appears the affine Goldstone boson serving as the graviton in
disguise. This results in  the effective Riemannian geometry with the
effective metric $g_{\mu\nu}$, etc. Now there appear the preferred
time and space directions, the lengths and angles,  the definite
Lorentz group and thus the finite dimensional spinors,
the definite  Poincare group and thus the particles, the invariant
intervals,  the quadratic invariants, the causality,~etc. 
The world structure becomes now very flourishing.
In the wake of the gravity, there appear the conventional 
matter. The spontaneous breaking of the affine symmetry to the
Poincare one reflects the appearance of the coherent particle
structure, among a lot of  a priori possible ones corresponding
to the various choices of the Poincare subgroup. 
Formally, the effective Riemannian geometry  is  to be  valid at all
the space-time intervals. Nevertheless, its accuracy  worsen when
diminishing the intervals, requiring more and more terms in the
decomposition over the ratio of the energy to the symmetry breaking
scale~$M_A$, as it should be for the effective theory.
Thus,  the scale~$M_A$ (or, rather, the Planck mass $M_{Pl}$) is a
kind of the inverse minimal length in the nature.

\subsection{The Universe}

Conceivably, the formation of the Universe is the result of the
actual transition between the two phases of the continuum. 
This transition is thus  {\it the ``Grand Bang''}, the origin  not
only of the  Universe but of the very space-time. At this stage, there
appears the world ``arrow of time'' as the reflection of the
spontaneous synchronization of the chaotic local times. The residual
dependence of the structure  of the Universe on the background
parameter-functions could result in the variety of the primordial
effects, such as the anisotropy, inhomogeneity, etc.
And what is more, there is conceivable the appearance (as well as 
disappearance and coalescence) of the various regions of the metric
phase inside the affinely connected  one (and v.v.). 
These regions  are to be associated with the multiple universes. One
of the latter ones happens to be ours.
Call the ensemble of the universes {\it the Metauniverse}.  
Within the concept of the Metauniverse,  there becomes sensible the
notion of the wave function of the Universe. Hopefully, this may
clarify the long-standing problem of the fine tuning 
of our Universe.\footnote{Cf., e.g., ref.~\cite{Davis}.}

\section{Conclusion}\label{Con}

To conclude, the theory proposed realizes consistently the approach to
gravity as the Goldstone phenomenon. It proceeds, in essence,
from the two basic symmetries: the global affine one and the general
covariance. The affine symmetry is the structure symmetry 
which defines the theory in the small. 
The general covariance is the bundling symmetry which terminates  the
a priory admissible local theories according to their ability 
to be prolongated onto the space-time.
The theory embodies the GR as the lowest approximation. 
Its distinction with the GR are twofold. At the effective level, the
given theory predicts the natural hierarchy of the conceivable GR
extensions, according to the accuracy of the affine symmetry
realization.  At the underlying level, the theory presents the new
look at the gravitation, the Universe  and the very space-time. 

The author is grateful to V.~V.~Kabachenko for useful discussions.

\end{document}